\documentclass[12pt,aaspp4]{aastex}

\usepackage{natbib}
\begin{document}
\newcommand{\up}[1]{\ifmmode^{\rm #1}\else$^{\rm #1}$\fi}
\newcommand{\zdot}{\makebox[0pt][l]{.}}
\newcommand{\upd}{\up{d}}
\newcommand{\uph}{\up{h}}
\newcommand{\upm}{\up{m}}
\newcommand{\ups}{\up{s}}
\newcommand{\arcd}{\ifmmode^{\circ}\else$^{\circ}$\fi}
\newcommand{\arcm}{\ifmmode{'}\else$'$\fi}
\newcommand{\arcs}{\ifmmode{''}\else$''$\fi}

\title{The Araucaria Project. The Distance of the Large Magellanic Cloud
from Near-Infrared Photometry of RR Lyrae Variables
\footnote{Based on observations obtained with the ESO NTT for 
programme 074.D-0318(B)}
}
\author{Olaf Szewczyk}
\affil{Universidad de Concepci{\'o}n, Departamento de Fisica, Astronomy
Group,
Casilla 160-C,
Concepci{\'o}n, Chile}
\affil{Warsaw University Observatory, Al. Ujazdowskie 4, 00-478, Warsaw,
Poland}
\authoremail{szewczyk@astro-udec.cl}
\author{Grzegorz Pietrzy{\'n}ski}
\affil{Universidad de Concepci{\'o}n, Departamento de Fisica, Astronomy
Group,
Casilla 160-C,
Concepci{\'o}n, Chile}
\affil{Warsaw University Observatory, Al. Ujazdowskie 4, 00-478, Warsaw,
Poland}
\authoremail{pietrzyn@astrouw.edu.pl}
\author{Wolfgang Gieren}
\affil{Universidad de Concepci{\'o}n, Departamento de Fisica, Astronomy Group,
Casilla 160-C, Concepci{\'o}n, Chile}
\authoremail{wgieren@astro-udec.cl}
\author{Jesper Storm}
\affil{Astrophysikalisches Institut Potsdam, An der Sternwarte 16,
D-14482
Potsdam, Germany}
\authoremail{jstorm@aip.de}
\author{Alistair Walker}
\affil{Cerro Tololo Inter-American Observatory, Casilla 603, La Serena,
Chile}
\authoremail{awalker@ctio.noao.edu}
\author{Luca Rizzi}
\affil{Joint Astronomy Centre, 66 N. A'ohoku Pl., Hilo, Hawaii, 96720}
\authoremail{luca@ifa.hawaii.edu}
\author{Karen Kinemuchi}
\affil{Universidad de Concepci{\'o}n, Departamento de Fisica, Astronomy
Group,
Casilla 160-C, Concepci{\'o}n, Chile}
\affil{University of Florida,
Department of
Astronomy, Gainesville, Florida 32611-2055, USA}
\authoremail{kkinemuchi@astro-udec.cl}
\author{Fabio Bresolin}
\affil{Institute for Astronomy, University of Hawaii at Manoa, 2680 Woodlawn 
Drive,  Honolulu HI 96822, USA}
\authoremail{bresolin@ifa.hawaii.edu}
\author{Rolf-Peter Kudritzki}
\affil{Institute for Astronomy, University of Hawaii at Manoa, 2680 Woodlawn 
Drive, Honolulu HI 96822, USA}
\authoremail{kud@ifa.hawaii.edu}
\author{Massimo Dall'Ora}
\affil{INAF, Osservatorio Astronomico di Capodimonte, I-80131 Napoli, Italy}
\authoremail{dallora@oacn.inaf.it}

\begin{abstract}
We have obtained deep infrared $J$ and $K$ band observations of five 
fields located in the Large Magellanic Cloud (LMC) bar with the ESO
New Technology Telescope equipped with the SOFI infrared camera. In
our fields, 65 RR Lyrae stars catalogued by the OGLE collaboration
were identified. Using different theoretical and empirical calibrations of the
period-luminosity-metallicity relation, we find consistent LMC
distance moduli values.  Since the observed fields are situated very
close to the center of the LMC, the correction for the tilt of the LMC bar with
respect to the line of sight is negligible. Our adopted best true
distance modulus to the LMC of $18.58 \pm 0.03$ (statistical) $\pm$ 
0.11 (systematic) mag agrees very well with most independent
determinations to this galaxy.
\end{abstract}


\section{Introduction}
In our ongoing Araucaria Project (e.g. Gieren et al. 2005a), we are applying a
number of different stellar standard candles to independently determine the
distances to a sample of nearby galaxies.  The systematic differences
between the distance results obtained for the individual galaxies from
the various stellar candles will be analyzed in forthcoming
papers. This analysis is expected to finally lead to a detailed
understanding of how the various stellar techniques, which are
fundamental to calibrate the first rungs of the distance ladder, depend
on metallicity and age.  While the objects we use for the
distance determinations are usually detected from optical wide-field imaging
surveys of the target galaxies (e.g. Pietrzy{\'n}ski et al. 2002b), the most
accurate distance work is then done from follow-up near-infrared images which
virtually eliminate reddening as a significant source of error on the
results. Examples of this very successful approach are the recent Cepheid work
on NGC 55 (Gieren et al. 2008), and the red clump star work on the LMC
(Pietrzy{\'n}ski \& Gieren 2002a). The Araucaria Project has also been
developing a new spectroscopic distance indicator, viz. the flux-weighted
gravity-luminosity relationship for blue supergiants (Kudritzki et al. 2003;
2008) which holds the promise to yield distances accurate to 5\% to galaxies
containing massive blue stars out to about 10 Mpc from low-resolution spectra.

Thanks to several recent theoretical and empirical studies, evidence has been
mounting that RR Lyrae (RRL) stars are excellent standard candles in the
near-infrared spectral range, providing distance results which are superior to
the traditional optical method (e.g. Bono 2003a).  Longmore et al. (1986) were
the first to show that RRL variable stars follow a period-luminosity (PL)
relation in the near-infrared K-band.  Their pioneering work was followed by
Liu \& Janes (1990), 
Jones et al. (1996),
%
and Skillen et al. (1993),
who applied infrared versions of the Baade-Wesselink method to
calibrate the luminosities and distances of RRL stars.  A very
comprehensive analysis of the IR properties of RRL stars was
given by Nemec et al. (1994).  The first theoretical constraints on the
K-band PL relation of RRL stars are based on non-linear convective
pulsation models that were presented by Bono et al. (2001).  Dall'Ora et
al. (2004) later demonstrated that the K-band PL relation for RRL
stars appears to have a very small scatter for globular clusters, with 
small intrinsic spread in metallicity for this type of stars. Further theoretical
explorations of the RRL period-mean magnitude-metallicity
relations in near-infrared passbands were carried out by Bono et
al. (2003b), Catelan et al. (2004), and Cassisi et al. (2004). Most recently, Sollima et
al. (2006) analyzed near-infrared K-band data of RRL stars in
some 15 Galactic globular clusters and provided the first empirical
calibration of the period-luminosity-metallicity (PLZ) relation in the
$K$ band. All these existing theoretical and empirical studies have
suggested that the RRL star K-band PLZ relation appears to be
indeed a superb means to determine accurate distances to galaxies
hosting an abundant old stellar population. 

We have therefore started to include this rather new tool in the
Araucaria Project distance work.  Pietrzy{\'n}ski et al. (2008)
implemented the RRL K-band PLZ relation for single-epoch K-band
magnitudes of a large sample of RRL stars in the Sculptor dwarf
galaxy.  The distance determination from their analysis for this
galaxy compares very well with the distance of Sculptor determined
from the tip of the red giant branch method.  In this paper, we are
applying the RRL K-band PLZ relation to a sample of RRL stars in five fields
of the Large Magellanic Cloud (LMC). 
This is especially important
because the LMC Cepheid PL relation has served as the fiducial relation
to which the Cepheid distances measured to a sample of late-type
galaxies by the two HST Key Projects (Freedman et al. 2001; Saha et al.
2001) has been tied.
On the other
hand, there is still an uncomfortably large and annoying discrepancy
among modern distance determinations to the LMC from different methods
(e.g. Schaefer 2008; Gieren et al. 2005b; Walker 2003; Feast 2003;
Benedict el al. 2002). Most recent distance determinations to the LMC
have clustered around the value of 18.5 mag that was adopted by the Key Project
(but see a cautionary comment made by Schaefer (2008)).  Discrepant results
continue to be obtained, as shown by recent exhaustive determinations of Milky
Way Cepheid distances from the infrared surface brightness technique
(Fouqu{\'e} et al. 2007; Fouqu{\'e} \& Gieren 1997). A shorter
LMC distance modulus (closer to 18.4 mag) was suggested by this method.
A first determination of the LMC distance from an application of the
K-band PLZ relation on the LMC field RRL variable stars is an
important step towards a resolution of the distance discrepancy.  
This is urgently needed for true progress on the calibration of the
distance scale.

\section{Observations, Data Reduction and Calibration}

All the near-infrared data used and presented in this paper were collected
during one of the Araucaria Project observing runs. The SOFI infrared camera
of the ESO New Technology Telescope (NTT) telescope at La Silla
Observatory was used. With the Large Field setup we achieved a $4.9 \times
4.9$ arcminute field of view and a pixel scale of 0.288 arcsec/pixel. 

During two photometric nights we obtained deep $Js$ and $Ks$ observations of
five fields in the LMC, with each field containing at least a dozen RRL stars.
Fig. \ref{figfields} displays the location of these fields in the
LMC. On the second night, fields 1 and 2 were overlapped with field
3a. Detailed information on each field is given in Table \ref{tabfields}. 
In order to take into consideration the rapid sky level variations in the
infrared passband, we used a dithering technique. Total integration
times were up to 40 minutes for $Ks$, and 11 minutes for the $Js$ band. 

The pipeline developed in the course of Araucaria Project was used for all the
reductions and calibrations. First, the subtraction of sky level was applied
in a two-step process which includes the masking of stars with the IRAF xdimsum
package (Pietrzy{\'n}ski \& Gieren 2002a). Next, each single image was
flat fielded and stacked into the final deep field. PSF photometry, including
aperture corrections, was performed in the same way as described in
Pietrzy{\'n}ski, Gieren \& Udalski (2002c). 

The calibration of the photometry onto the standard system was based on
observations of 14 standard stars from the UKIRT list (Hawarden et
al. 2001). All of them were observed together with the target fields during
photometric conditions at different airmasses.  Thanks to the large number of
standard stars observed along with the science target fields, the accuracy of
our photometry zero point was estimated to be as good as 0.02 mag.
Our calibrated photometric magnitudes were compared with the 2MASS catalogue
for common stars, which gave us a zero point difference.  Also stars that were
measured on both nights and cross-identified on different fields were
compared (SC5-FI + SC5-FII with SC5-FIII, and SC7-FV, on following
nights). 
The results of comparison of zero point differences between our work and 2MASS are shown in Table \ref{tab2mass}.  
%
The increase in difference between K-band observations for field 3a may be caused by larger crowding and less accuracy of 2MASS photometry in these regions.
The Red Clump (RC) brightness was
also compared with previously published data (Pietrzy{\'n}ski,
Gieren \& Udalski 2003). Results of the determination of the RC star mean
brightness in each of our fields are given in Tab. \ref{tabrc}. They compare very well
with the values found by Pietrzy{\'n}ski \& Gieren (2002a) in their observed
fields ($J=17.507 \pm 0.009$, $K=16.895 \pm 0.007$).  The calibrated
near-infrared magnitudes for all RRL stars identified in our
fields are presented in Tab. \ref{tabrawobs}.

\section{Near-Infrared Period-Luminosity Relations}

The sample of 65 RRL stars we have observed in our chosen SOFI/NTT fields
were cross-identified with the OGLE catalogue of RRL stars in the LMC
(Soszy{\'n}ski et al. 2003). The position of all identified stars in the $K$, $J-K$
color-magnitude diagram is shown on Fig. \ref{figcmd}. 
Most of the stars have only one random-phase measurement, but some
have two random-phase observations taken on the two different nights. A
few of the RRL stars were cross-identified in the overlapping
fields, which provided three measurements for these objects. 

The period-luminosity (PL) relations for the $J$ and $K$ bands derived from our
data are shown in Fig. \ref{figplav}. For the RRL stars with more than one
observation, we took a straight average of the random-phase magnitudes, which
should lead to a better approximation of their mean magnitudes.  In
Fig. \ref{figplav}, two distinct groups can be distinguished that correspond to the
first overtone (RRc) and fundamental mode (RRab) pulsators,
respectively. The relatively large scatters seen in both figures \ref{figcmd} and \ref{figplav} is
mostly caused by three factors: 1) the random single-phase nature of
our IR measurements, which represents the mean magnitude of an RRL variable
only to $\sim 0.15$ mag (e.g. Del Principe et al. 2006), 2)
the metallicity spread among RRL stars in the LMC, and 3) the accuracy of our
single measurements, which is 0.03-0.19 mag for stars of brightnesses of
16.6-18.6 mag in the $K$ band. 

Since the most important contributor to the scatter in Figure \ref{figplav} is
the replacement of the mean magnitudes by single-phase or by averaged
few-phase measurements, we used accurate optical ($BVI$) photometry
from the OGLE archive (Soszy{\'n}ski et al. 2003) of the the studied sample
of RRL stars, taken close in time to the NIR data.  With the additional
optical data, we can calculate improved $JK$ phase points in
order to use the 
Jones et al. (1996)
%
template light curve method
(available only for the $K$ band).
The $J$, $K$, and $\langle K \rangle$ magnitudes are given in Table \ref{tabavobs}.

 We compare the PL relations derived for the RRab stars (averaged
 observations) and for all of the RRL stars (RRab and RRc) against
 the existing theoretical (Bono et al. 2003b; Catelan et al. 2004) and
 empirical (Sollima et al. 2006; 2008) relations.  Table \ref{tabplav} lists our results for the
 slope and zero point values of the PL relations as well as those
 from the theoretical and empirical ones.  Furthermore, in Table
 \ref{tabplmean}, we present the K-band zero point and slope values obtained from 
$\langle K \rangle$
 magnitudes derived from the 
Jones et al. (1996)
%
light curve template method.

\section{The Distance Determination}

In order to derive the apparent distance moduli to LMC from our data, we used
the following calibrations of the near-infrared PL relations of mixed
population RRL stars: 

 \begin{equation} M_{K} = -1.07 - 2.38\log P + 0.08[Fe/H]
 \qquad\mbox{-- Sollima et al. (2008)}
 \end{equation}

 \begin{equation} M_{K} = -0.77 - 2.101\log P + 0.231[Fe/H]
 \qquad\mbox{-- Bono et al. (2003b)}
 \end{equation}

 \begin{equation} M_{K} = -0.597 - 2.353\log P + 0.175\log Z
 \qquad\mbox{-- Catelan et al. (2004)}
 \end{equation}

 \begin{equation} M_{J} = -0.141 - 1.773\log P + 0.190\log Z
 \qquad\mbox{-- Catelan et al. (2004)}
 \end{equation}

We recall that the calibration of Sollima et al. (2008) was constructed for the 2MASS
photometric system, while the calibrations of Catelan et al. (2004)  and Bono et
al. (2003b) are valid for the Glass and Bessel and Brett systems,
respectively. Therefore, we transformed our own data, calibrated onto the
UKIRT system (Hawarden et al. 2001) to the Glass and Bessel and Brett  systems
using the transformations given by Carpenter (2001) before calculating
distances using the calibrations of Catelan et al. (2004) and Bono et
al. (2003b).  Since there is virtually no difference between the K-band
of 2MASS and UKIRT (Carpenter 2001), we did not apply any
transformations to our data while using the Sollima et al. (2008) calibration. 

In order to combine the RRab and RRc stars, we fundamentalize the RRc periods by adding $\log P = 0.127$.
Assuming the mean metallicity of our RRL sample to be $[Fe/H] = -1.48$ (Gratton et al. 2004), we have
calculated the $K$ and $J$ band distance moduli for the averaged magnitude 
and mean magnitude based on the template light-curve data. The fits for
the relations 1-4 to both sets of data are displayed in Figs \ref{figfitjk} and \ref{figfitk}. 

To correct the derived apparent distance moduli for interstellar reddening,
we adopted reddening maps of $E(B-V)$ calculated for the LMC by Udalski et
al. (1999). Assuming the reddening law from 
Fitzpatrick (1999),
%
we calculate the following selective extinctions in the
different bands: $A_{K} = 0.367E(B-V)$ and $A_{J}= 0.902E(B-V)$. 

For the averaged $K$ band data the true distance moduli of $18.58 \pm
0.03$, $18.62 \pm 0.03$, and $18.60 \pm 0.03$ mag, were obtained using the
calibrations of Sollima et al. (2008), Bono et al (2003b) and Catelan et
al. (2004), respectively. Based on the $J$ band data and the calibration of
Catelan et al. (2004), the distance moduli of $18.55 \pm 0.03$ mag was
derived. Similar values of the true distance moduli were calculated using the
mean 
$\langle K \rangle$ 
band magnitudes obtained from the 
Jones et al. (1996)
%
template fitting
($18.56  \pm 0.03$ - Sollima et al. (2008), $18.60 \pm 0.03$ - Bono et al. (2003b),
$18.59 \pm 0.03$ - Catelan et al. (2004)). These results are summarized in
Tab. \ref{tabdist}.  For our best distance determination, we adopt the average
value from all the results to be of $18.58 \pm 0.03$ mag. 

\section{Discussion}

The distance moduli obtained based on several independent theoretical
and empirical calibrations are consistent. The maximum difference 
of 0.04 mag between the results from the calibrations of Sollima et al.
(2008) and Bono et al. (2003b) is certainly not significant taking into 
account all uncertainties, which affect the whole process of
constructing the mentioned calibrations. However, it is interesting to note 
that a very similar difference between the distance moduli derived using
these two calibrations was recently obtained by Pietrzy{\'n}ski et al. 
(2008) for the Sculptor galaxy ($[Fe/H] = -1.83$ dex). Therefore,
perhaps there is just a zero point offset in the sense that the
distances from the  calibration of Sollima et al. (2008) are slightly
shorter compared to those from the calibration of Bono et al. (2003b).

Taking into account the errors associated with the adopted calibrations,
mean metallicity, photometric zero point and absorption correction,
we estimate the systematic error of our distance determination to be of
0.11 mag. Therefore our best distance modulus determination to the LMC
is: $18.58 \pm 0.03$ (statistical) $\pm 0.11$ (systematic) mag.

It is worthwhile to mention that the observed fields are located not only close 
to the LMC center, but also opposite each other around it,
so the corrections for the tilt of this galaxy with 
respect to the line of sight are expected to be very small. 
Indeed applying the geometrical model of van der Marel et al. 
(2002) to correct our data for this effect, we obtain a distance
modulus shorter by 0.01 mag. 
    Comparing our distance result from the present field RR Lyrae stars to
    the distance obtained by Dall'Ora el al. (2004) for the Reticulum
    cluster, there is evidence that the cluster could be very slightly
    nearer than the LMC center, by about 3\%, but this small difference
    is clearly within the combined uncertainties, even the statistical
    ones, of both determinations.

Very recently Sollima et al. (2008), using their calibration and 
the IR data of a sample of RRL stars presented by Borissova
et al. (2004), obtained a distance modulus to the LMC of $18.56 \pm 
0.13$ mag. Our distance moduli derived based on this same calibration are 
virtually the same, which reinforces both results.

Our distance modulus agrees very well with the most LMC distance moduli 
derived from other independent techniques (Freedman et al. 2001;
Benedict et al. 2002; Walker 2003). In particular this result is very
similar to the measurements obtained based on the near infrared
photometry of Cepheids (Persson et al. 2004) and the red clump stars
(Alves et al. 2002; Grocholski \& Sarajedini 2002; Pietrzy{\'n}ski \& 
Gieren 2002a). 

\section{Summary and Conclusions}
The results of our deep infrared imaging of 65 RR Lyrae stars in the central
regions of the LMC are presented.  Our data shows two clear sequences
in the period luminosity plane, which correspond to the RRc and RRab
stars. After fundamentalizing the RRc periods to the period of
the RRab stars by adding $\log P = 0.127$, both groups were
merged, and the distance moduli to the LMC were determined using
different theoretical and empirical calibrations.  Our final adopted
distance agrees very well with the recent results obtained from the
near infrared observations of Cepheids and red clump stars as well as
with those calculated from other techniques. 

Our results confirm that the RR Lyrae period-luminosity-metallicity 
relations in the near infrared passband are potentially a very good tool 
for precise distance measurements.

\acknowledgments
WG and  GP  gratefully acknowledge financial support for this
work from the Chilean Center for Astrophysics FONDAP 15010003. 
Support from the Polish grant N203 002 31/046 and the FOCUS
subsidy of the Fundation for Polish Science (FNP)
is also acknowledged.
WG gratefully acknowledges support for this work from the Chilean
Centro de Astrofisica y Tecnologias Afines, CATA.
It is a special pleasure to thank the support astronomers at ESO-La Silla 
for their expert help in the observations, and the ESO OPC for the
generous amounts of observing time at the NTT allocated to our programme.
We thank the referee, Dr. Giuseppe Bono, for constructive remarks
which helped to improve the paper.

\setcounter{figure}{0}

\clearpage
\begin{figure}[p] 
\vspace*{18cm}
\includegraphics{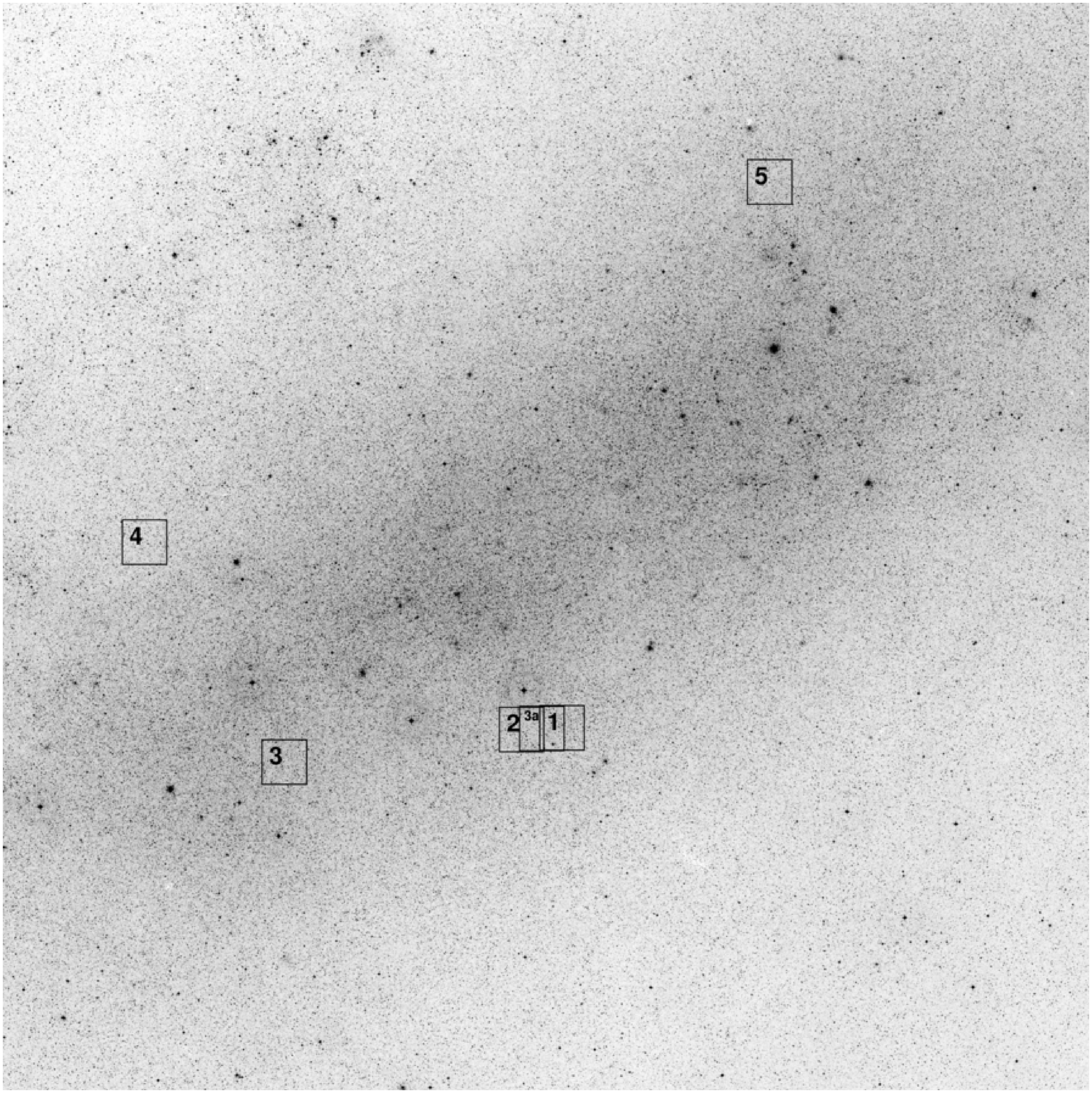} 
\caption{The location of our observed $5 \times 5$ arcmin  NTT/SOFI fields 
in LMC on the DSS-2 infrared plate. North is up and east to the left. 
}
\label{figfields}
\end{figure}  

\clearpage
\begin{figure}[htb]
\vspace*{15cm}
\includegraphics{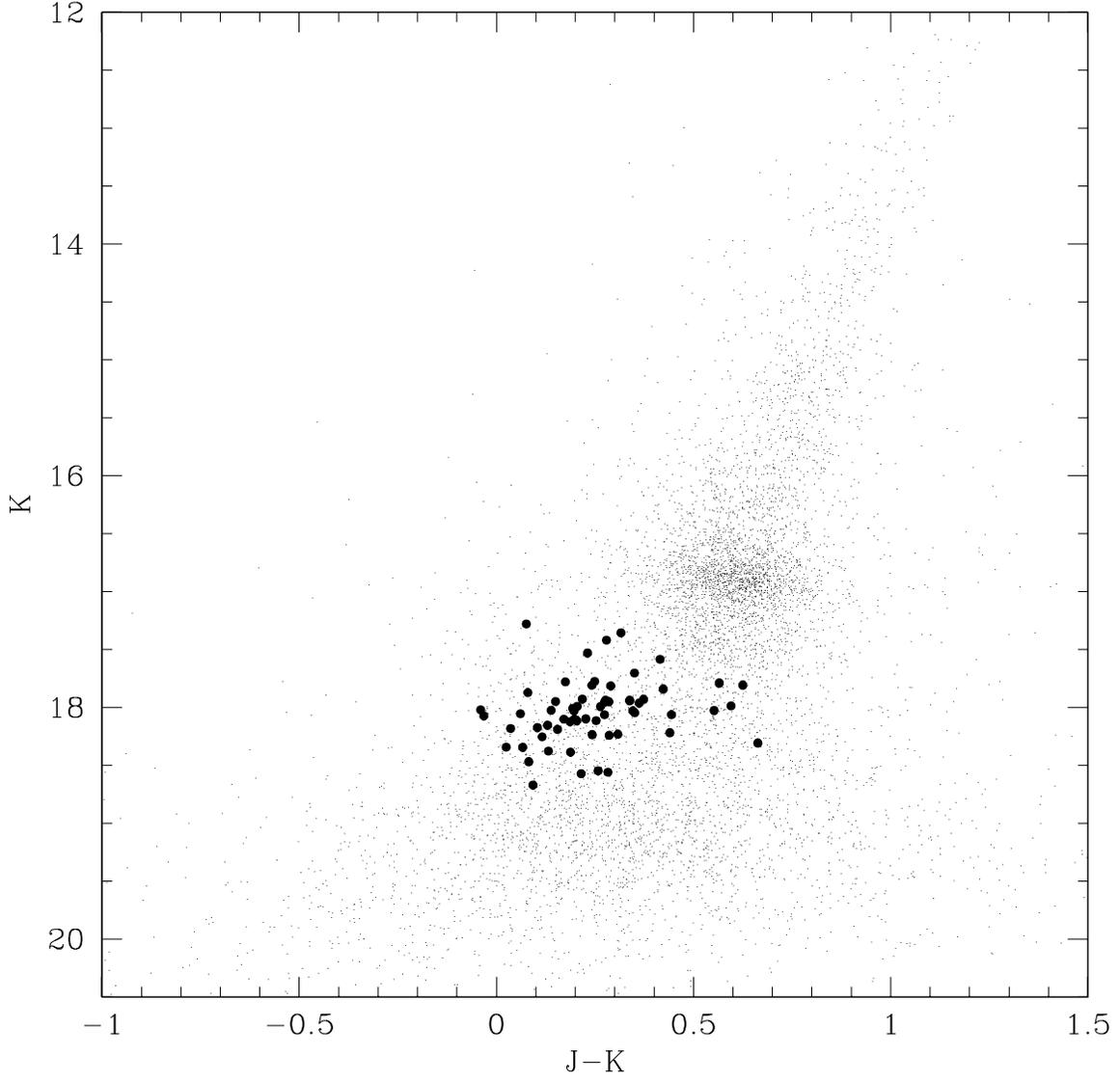} 
\caption{The color-magnitude diagram showing identified RR Lyrae stars
  (filled circles) in observed fields . 
}
\label{figcmd}
\end{figure}  

\clearpage
\begin{figure}[htb]
\vspace*{15cm}
\includegraphics{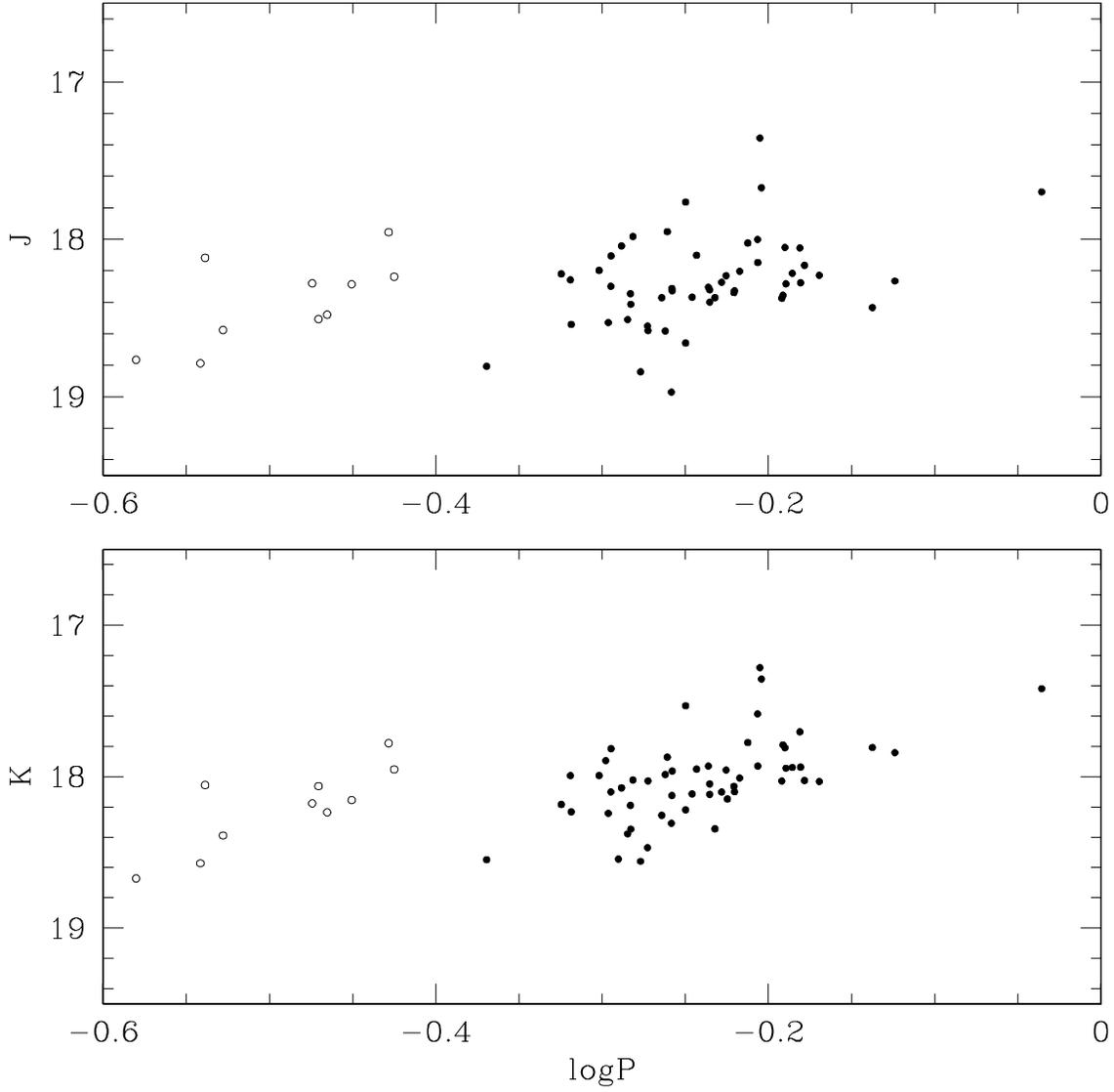}
\caption{The near-infrared K and J band period-luminosity relations 
defined by the 65 RR Lyrae stars observed in LMC. Period is in days.
Two distinct groups are formed by the fundamental (filled circles) and first overtone (open circles)
pulsators which are clearly seen in each panel.}
\label{figplav}
\end{figure}

\clearpage
\begin{figure}[htb]
\vspace*{15cm}
\includegraphics{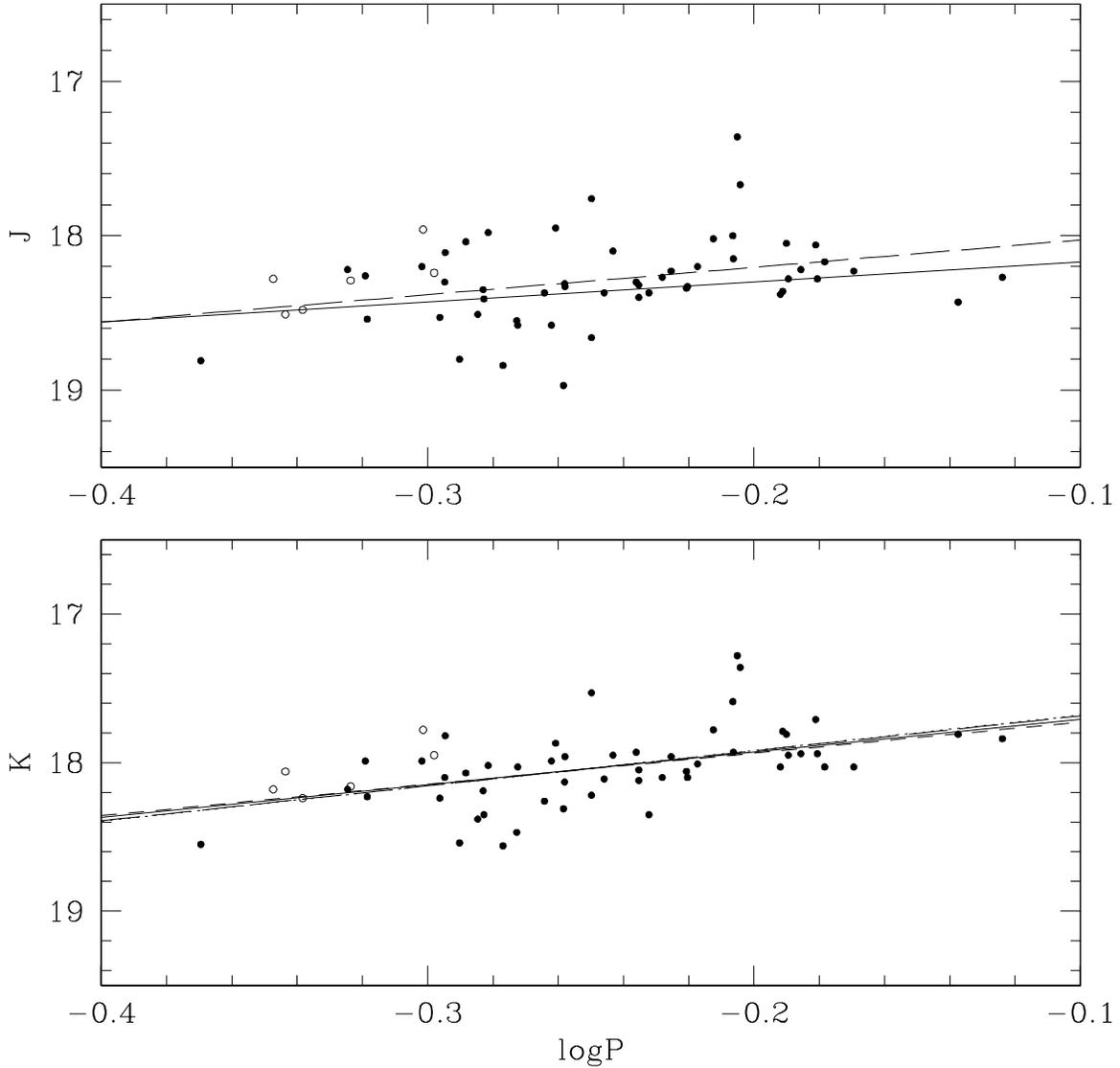}
\caption{
The near-infrared PL relations in $K$ and $J$ defined by our RR Lyrae
sample in LMC, plotted along with the best-fitting lines. The slopes
of the fits were adopted from the recent theoretical and empirical
calibrations, and the zero points determined from our data. The solid,
dotted, short and long dashed lines correspond to free fit and the
calibration of Sollima et al. (2008), Bono et al. (2003), and Catelan et al. (2004),
respectively.}
\label{figfitjk}
\end{figure}

\clearpage
\begin{figure}[htb]
\vspace*{8cm}
\includegraphics{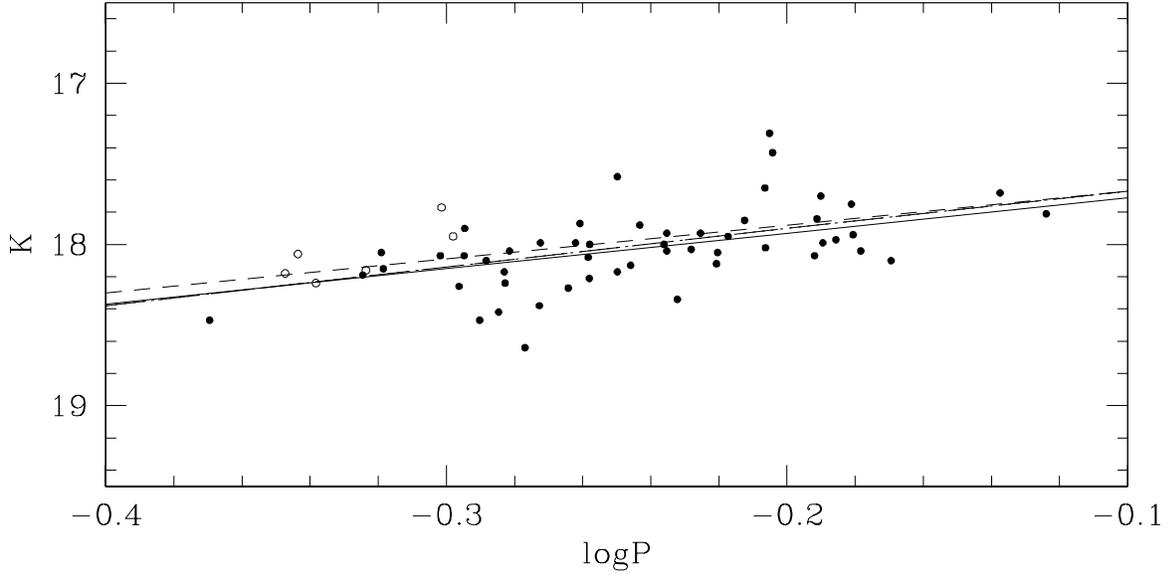}
\caption{The near-infrared PL relations in 
$\langle K \rangle$ 
band (corrected with
  light-curve templates) defined by our RR Lyrae sample in LMC,
  plotted along with the best-fitting lines. The slopes of the fits
  were adopted from the recent theoretical and empirical calibrations,
  and the zero points determined from our data. The solid, dotted,
  short and long dashed lines correspond to free fit and the
  calibration of Sollima et al. (2008), Bono et al. (2003), and Catelan et al. (2004),
  respectively.} 
\label{figfitk}
\end{figure}

\setcounter{table}{0}
\clearpage
\begin{deluxetable}{c c c c c c}
\tablewidth{0pc}
\tablecaption{Observational information of the target fields.
Extinction values based on reddening maps by Udalski et al. (1999).
}
\tablehead{ \colhead {Field} & \colhead{Field} & \colhead {RA2000} & \colhead {DEC2000} & \colhead {Date of} & \colhead {Extinction} \\
\colhead {No} & \colhead {name} & \colhead {} & \colhead {} & \colhead {observation} & \colhead {E(B-V)} }
\startdata
1   &  SC5-FI    &      05:23:24.0   &   -70:05:24.0 &    2006-09-22             & 0.130 \\
2   &  SC5-FII   &      05:24:16.7   &   -70:05:24.0 &    2006-09-22             & 0.130 \\
3a  &  SC5-FIII  &      05:23:50.3   &   -70:05:24.0 &    2007-04-06             & 0.130 \\
3   &  SC3-FIII  &      05:29:26.3   &   -70:07:30.0 &    2007-04-06             & 0.134 \\
4   &  SC2-FIV   &      05:32:05.0   &   -69:42:12.9 &    2007-04-06             & 0.150 \\
5   &  SC7-FV    &      05:18:38.4   &   -69:06:00.0 &    2006-09-22, 2007-04-06 & 0.146 \\

\enddata
\label{tabfields}
\end{deluxetable}

\clearpage
\begin{deluxetable}{c c c c}
\tablewidth{0pc}
\tablecaption{Difference of zeropoint estimation between 2MASS and observed data.}
\tablehead{\colhead {Field No} &\colhead{Field name}&\colhead{2MASS-J20070406}&\colhead{2MASS-K20070406}}
\startdata
3a & SC5-FIII     &  0.03 $\pm$ 0.13 &  0.13 $\pm$ 0.30 \\
3  & SC3-FIII     &  0.00 $\pm$ 0.10 &  0.02 $\pm$ 0.10 \\
4  & SC2-FIV      &  0.09 $\pm$ 0.15 &  0.03 $\pm$ 0.11 \\
5  & SC7-FV       &  0.09 $\pm$ 0.13 &  0.01 $\pm$ 0.11 \\
\enddata
\label{tab2mass}
\end{deluxetable}

\clearpage
\begin{deluxetable}{c c c c}
\tablewidth{0pc}
\tablecaption{Comparison of red clump star brightnesses for the observed fields.}
\tablehead{ \colhead {Field} & \colhead {Date} & \colhead {J RC} & \colhead {K RC} }
\startdata
1  SC5-FI    &   2006-09-22    &  17.358$\pm$0.011     &      16.895$\pm$0.006 \\
2  SC5-FII   &   2006-09-22    &  17.504$\pm$0.006     &      16.950$\pm$0.004 \\
3a SC5-FIII  &   2007-04-06    &  17.437$\pm$0.009     &      16.917$\pm$0.007 \\
3  SC3-FIII  &   2007-04-06    &  17.492$\pm$0.014     &      16.900$\pm$0.007 \\
4  SC2-FIV   &   2007-04-06    &  17.459$\pm$0.012     &      16.924$\pm$0.006 \\
5  SC7-FV    &   2006-09-22    &  17.510$\pm$0.009     &      16.956$\pm$0.007 \\
5  SC7-FV    &   2007-04-06    &  17.505$\pm$0.012     &      16.950$\pm$0.008 \\
\enddata
\label{tabrc}
\end{deluxetable}

\clearpage
\begin{deluxetable}{ccccccccc}
\rotate
\tablecaption{Individual J and K band Observations of RR Lyrae stars in LMC fields}
\tablewidth{0pc}
\tablehead{\colhead{Star} & \colhead{Star} & \colhead{Field} & \colhead{J HJD} & \colhead{J mag} & \colhead{$\sigma$} & \colhead{K HJD} & \colhead{K mag} & \colhead{$\sigma$} \\
\colhead{ID} & \colhead{type} & \colhead{name} & \colhead{+2400000} & \colhead{} & \colhead{} & \colhead{+2400000} & \colhead{} & \colhead{} }
\startdata
OGLE051817.95-690358.9	& ab	& LMC-SC7-FV	& 54001.30645	& 18.84	& 0.05	& 54001.31523	& 18.44	& 0.06 \\
			& 	& 		& 54196.98138	& 18.78	& 0.06	& 54196.98915	& 18.66	& 0.12 \\
OGLE051818.33-690513.2	& c	& LMC-SC7-FV	& 54001.30645	& 18.31	& 0.04	& 54001.31523	& 18.17	& 0.05 \\
			& 	& 		& 54196.98138	& 18.26	& 0.07	& 54196.98915	& 18.14	& 0.07 \\
OGLE051818.71-690506.2	& ab	& LMC-SC7-FV	& 54001.30645	& 17.73	& 0.03	& 54001.31523	& 17.42	& 0.03 \\
			& 	& 		& 54196.98138	& 17.61	& 0.04	& 54196.98915	& 17.30	& 0.05 \\
OGLE051820.31-690819.4	& c	& LMC-SC7-FV	& 54001.30645	& 18.10	& 0.06	& 54001.31523	& 18.10	& 0.05 \\
			& 	& 		& 54196.98138	& 18.13	& 0.06	& 54196.98915	& 18.02	& 0.08 \\
OGLE051824.42-690459.4	& ab	& LMC-SC7-FV	& 54001.30645	& 18.40	& 0.04	& 54001.31523	& 18.08	& 0.04 \\
			& 	& 		& 54196.98138	& 18.17	& 0.04	& 54196.98915	& 17.81	& 0.06 \\
OGLE051834.72-690550.6	& ab	& LMC-SC7-FV	& 54001.30645	& 18.15	& 0.04	& 54001.31523	& 17.97	& 0.04 \\
			& 	& 		& 54196.98138	& 18.25	& 0.04	& 54196.98915	& 18.02	& 0.07 \\
OGLE051835.56-690406.0	& ab	& LMC-SC7-FV	& 54001.30645	& 18.20	& 0.05	& 54001.31523	& 18.09	& 0.05 \\
			& 	& 		& 54196.98138	& 18.24	& 0.06	& 54196.98915	& 18.28	& 0.08 \\
OGLE051836.59-690839.6	& ab	& LMC-SC7-FV	& 54001.30645	& 18.55	& 0.09	& 54001.31523	& 18.47	& 0.05 \\
OGLE051836.68-690659.7	& ab	& LMC-SC7-FV	& 54001.30645	& 18.27	& 0.07	& 54001.31523	& 18.10	& 0.04 \\
OGLE051837.86-690821.3	& c	& LMC-SC7-FV	& 54001.30645	& 18.80	& 0.07	& 54001.31523	& 18.57	& 0.06 \\
			& 	& 		& 54196.98138	& 18.77	& 0.06	& 54196.98915	& 18.57	& 0.10 \\
OGLE051841.62-690609.4	& ab	& LMC-SC7-FV	& 54001.30645	& 18.36	& 0.03	& 54001.31523	& 17.87	& 0.04 \\
			& 	& 		& 54196.98138	& 18.17	& 0.04	& 54196.98915	& 17.81	& 0.06 \\
OGLE051842.24-690724.5	& ab	& LMC-SC7-FV	& 54001.30645	& 18.28	& 0.03	& 54001.31523	& 17.94	& 0.03 \\
OGLE051843.33-690502.6	& ab	& LMC-SC7-FV	& 54001.30645	& 18.43	& 0.05	& 54001.31523	& 18.14	& 0.05 \\
			& 	& 		& 54196.98138	& 18.31	& 0.05	& 54196.98915	& 18.09	& 0.07 \\
OGLE051853.07-690819.5	& ab	& LMC-SC7-FV	& 54196.98138	& 18.30	& 0.04	& 54196.98915	& 17.93	& 0.06 \\
OGLE051923.06-693859.0	& ab	& LMC-SC7-FV	& 54001.30645	& 18.36	& 0.05	& 54001.30645	& 17.80	& 0.07 \\
			& 	& 		& 54196.98138	& 18.35	& 0.07	& 54196.98915	& 17.79	& 0.07 \\
OGLE052301.40-700731.0	& c	& LMC-SC5-FI	& 54001.21410	& 18.58	& 0.05	& 54001.22293	& 18.39	& 0.05 \\
OGLE052308.89-700746.7	& c	& LMC-SC5-FI	& 54001.21410	& 18.48	& 0.05	& 54001.22293	& 18.24	& 0.05 \\
OGLE052317.13-700607.2	& ab	& LMC-SC5-FI	& 54001.21410	& 18.33	& 0.08	& 54001.22293	& 18.10	& 0.08 \\
OGLE052318.43-700651.5	& ab	& LMC-SC5-FI	& 54001.21410	& 18.33	& 0.05	& 54001.22293	& 17.96	& 0.04 \\
OGLE052321.67-700739.9	& ab	& LMC-SC5-FI	& 54001.21410	& 18.41	& 0.06	& 54001.22293	& 18.35	& 0.04 \\
OGLE052328.70-700523.8	& c	& LMC-SC5-FI	& 54001.21410	& 18.13	& 0.08	& 54001.22293	& 17.83	& 0.07 \\
			& 	& LMC-SC5-FIII	& 54196.99858	& 17.78	& 0.05	& 54197.00711	& 17.73	& 0.07 \\
OGLE052329.54-700713.1	& ab	& LMC-SC5-FI	& 54001.21410	& 17.90	& 0.04	& 54001.22293	& 17.66	& 0.03 \\
			& 	& LMC-SC5-FIII	& 54196.99858	& 18.21	& 0.08	& 54197.00711	& 17.75	& 0.05 \\
OGLE052331.40-700529.1	& ab	& LMC-SC5-FI	& 54001.21410	& 17.92	& 0.06	& 54001.22293	& 17.79	& 0.04 \\
			& 	& LMC-SC5-FIII	& 54196.99858	& 18.12	& 0.07	& 54197.00711	& 17.76	& 0.08 \\
OGLE052332.28-700654.1	& ab	& LMC-SC5-FI	& 54001.21410	& 17.35	& 0.03	& 54001.22293	& 17.28	& 0.03 \\
			& 	& LMC-SC5-FIII	& 54196.99858	& 17.37	& 0.05	& 54197.00711	& 17.29	& 0.05 \\
OGLE052335.14-700505.4	& ab	& LMC-SC5-FI	& 54001.21410	& 18.21	& 0.05	& 54001.22293	& 18.06	& 0.04 \\
			& 	& LMC-SC5-FIII	& 54196.99858	& 18.39	& 0.05	& 54197.00711	& 18.14	& 0.08 \\
OGLE052340.40-700639.8	& ab	& LMC-SC5-FIII	& 54196.99858	& 18.04	& 0.07	& 54197.00711	& 18.07	& 0.08 \\
OGLE052341.81-700632.4	& ab	& LMC-SC5-FI	& 54001.21410	& 18.97	& 0.11	& 54001.22293	& 18.80	& 0.07 \\
			& 	& LMC-SC5-FIII	& 54196.99858	& 18.69	& 0.05	& 54197.00711	& 18.32	& 0.09 \\
OGLE052345.86-700504.2	& ab	& LMC-SC5-FI	& 54001.21410	& 17.85	& 0.05	& 54001.22293	& 17.79	& 0.06 \\
			& 	& LMC-SC5-FIII	& 54196.99858	& 18.35	& 0.05	& 54197.00711	& 18.11	& 0.08 \\
OGLE052349.66-700327.1	& ab	& LMC-SC5-FI	& 54001.21410	& 18.08	& 0.07	& 54001.22293	& 17.83	& 0.06 \\
			& 	& LMC-SC5-FII	& 54001.26192	& 18.15	& 0.07	& 54001.27070	& 18.23	& 0.07 \\
			& 	& LMC-SC5-FIII	& 54196.99858	& 18.27	& 0.07	& 54197.00711	& 18.02	& 0.08 \\
OGLE052350.37-700443.2	& ab	& LMC-SC5-FI	& 54001.21410	& 18.50	& 0.07	& 54001.22293	& 18.22	& 0.06 \\
			& 	& LMC-SC5-FII	& 54001.26192	& 18.61	& 0.05	& 54001.27070	& 18.26	& 0.05 \\
			& 	& LMC-SC5-FIII	& 54196.99858	& 18.48	& 0.05	& 54197.00711	& 18.25	& 0.09 \\
OGLE052353.56-700432.4	& ab	& LMC-SC5-FII	& 54001.26192	& 18.44	& 0.04	& 54001.27070	& 18.22	& 0.04 \\
			& 	& LMC-SC5-FIII	& 54196.99858	& 18.18	& 0.05	& 54197.00711	& 18.03	& 0.07 \\
OGLE052355.34-700802.6	& c	& LMC-SC5-FII	& 54001.26192	& 18.82	& 0.08	& 54001.27070	& 18.54	& 0.06 \\
			& 	& LMC-SC5-FIII	& 54196.99858	& 18.71	& 0.08	& 54197.00711	& 18.81	& 0.18 \\
OGLE052357.93-700322.9	& ab	& LMC-SC5-FII	& 54001.26192	& ...	& ...	& 54001.27070	& 18.56	& ... \\
			& 	& LMC-SC5-FIII	& 54196.99858	& 18.80	& 0.08	& 54197.00711	& 18.53	& 0.12 \\
OGLE052400.52-700520.9	& ab	& LMC-SC5-FII	& 54001.26192	& 18.40	& 0.03	& 54001.27070	& 17.98	& 0.04 \\
			& 	& LMC-SC5-FIII	& 54196.99858	& 18.40	& 0.04	& 54197.00711	& 18.12	& 0.07 \\
OGLE052405.11-700611.5	& ab	& LMC-SC5-FII	& 54001.26192	& 18.18	& 0.04	& 54001.27070	& 17.92	& 0.05 \\
			& 	& LMC-SC5-FIII	& 54196.99858	& 18.29	& 0.06	& 54197.00711	& 17.99	& 0.07 \\
OGLE052418.64-700307.7	& c	& LMC-SC5-FIII	& 54196.99858	& 18.28	& 0.11	& 54197.00711	& 18.18	& 0.16 \\
OGLE052419.91-700616.7	& ab	& LMC-SC5-FII	& 54001.26192	& ...	& ...	& 54001.27070	& 17.90	& ... \\
OGLE052423.93-700736.4	& ab	& LMC-SC5-FII	& 54001.26192	& 18.32	& 0.05	& 54001.27070	& 18.12	& 0.04 \\
OGLE052859.00-700822.2	& ab	& LMC-SC3-FIII	& 54197.01705	& 18.97	& 0.10	& 54197.02559	& 18.31	& 0.13 \\
OGLE052909.42-700828.1	& ab	& LMC-SC3-FIII	& 54197.01705	& 17.70	& 0.05	& 54197.02559	& 17.42	& 0.07 \\
OGLE052911.54-700836.5	& ab	& LMC-SC3-FIII	& 54197.01705	& 18.58	& 0.06	& 54197.02559	& 17.99	& 0.07 \\
OGLE052917.43-700734.5	& ab	& LMC-SC3-FIII	& 54197.01705	& 18.34	& 0.04	& 54197.02559	& 18.06	& 0.07 \\
OGLE052917.77-700535.6	& ab	& LMC-SC3-FIII	& 54197.01705	& 18.43	& 0.07	& 54197.02559	& 17.81	& 0.07 \\
OGLE052917.87-700841.1	& ab	& LMC-SC3-FIII	& 54197.01705	& 18.54	& 0.05	& 54197.02559	& 18.23	& 0.08 \\
OGLE052917.92-700712.2	& c	& LMC-SC3-FIII	& 54197.01705	& 18.51	& 0.07	& 54197.02559	& 18.06	& 0.10 \\
OGLE052929.39-700551.0	& ab	& LMC-SC3-FIII	& 54197.01705	& 18.00	& 0.04	& 54197.02559	& 17.59	& 0.06 \\
OGLE052932.88-700842.3	& ab	& LMC-SC3-FIII	& 54197.01705	& 18.37	& 0.06	& 54197.02559	& 18.26	& 0.09 \\
OGLE052933.54-700715.9	& ab	& LMC-SC3-FIII	& 54197.01705	& 18.58	& 0.05	& 54197.02559	& 18.03	& 0.08 \\
OGLE052936.94-700721.1	& ab	& LMC-SC3-FIII	& 54197.01705	& 18.37	& 0.05	& 54197.02559	& 18.35	& 0.08 \\
OGLE052937.91-700844.9	& c	& LMC-SC3-FIII	& 54197.01705	& 18.24	& 0.05	& 54197.02559	& 17.95	& 0.08 \\
OGLE052949.93-700558.2	& ab	& LMC-SC3-FIII	& 54197.01705	& 18.66	& 0.06	& 54197.02559	& 18.22	& 0.09 \\
OGLE053147.30-694349.6	& ab	& LMC-SC2-FIV	& 54197.04702	& 18.11	& 0.05	& 54197.05671	& 17.82	& 0.07 \\
OGLE053147.82-694142.2	& ab	& LMC-SC2-FIV	& 54197.04702	& 18.05	& 0.05	& 54197.05671	& 17.81	& 0.07 \\
OGLE053155.92-694232.4	& ab	& LMC-SC2-FIV	& 54197.04702	& 18.35	& 0.05	& 54197.05671	& 18.19	& 0.08 \\
OGLE053200.94-694219.1	& ab	& LMC-SC2-FIV	& 54197.04702	& 18.20	& 0.09	& 54197.05671	& 18.01	& 0.12 \\
OGLE053203.62-694210.7	& ab	& LMC-SC2-FIV	& 54197.04702	& 18.26	& 0.04	& 54197.05671	& 17.99	& 0.07 \\
OGLE053206.72-694342.1	& ab	& LMC-SC2-FIV	& 54197.04702	& 18.51	& 0.04	& 54197.05671	& 18.38	& 0.09 \\
OGLE053206.78-694224.7	& ab	& LMC-SC2-FIV	& 54197.04702	& 17.98	& 0.04	& 54197.05671	& 18.02	& 0.07 \\
OGLE053207.32-694117.2	& ab	& LMC-SC2-FIV	& 54197.04702	& 17.76	& 0.03	& 54197.05671	& 17.53	& 0.06 \\
OGLE053207.88-694058.7	& ab	& LMC-SC2-FIV	& 54197.04702	& 18.15	& 0.04	& 54197.05671	& 17.93	& 0.07 \\
OGLE053217.55-694317.6	& ab	& LMC-SC2-FIV	& 54197.04702	& 17.95	& 0.08	& 54197.05671	& 17.87	& 0.07 \\
OGLE053219.15-694215.2	& ab	& LMC-SC2-FIV	& 54197.04702	& 18.22	& 0.05	& 54197.05671	& 17.94	& 0.07 \\
OGLE053222.39-693952.9	& ab	& LMC-SC2-FIV	& 54197.04702	& 18.23	& 0.07	& 54197.05671	& 18.03	& 0.11 \\
OGLE053225.42-694438.2	& ab	& LMC-SC2-FIV	& 54197.04702	& 18.38	& 0.07	& 54197.05671	& 18.03	& 0.12 \\
OGLE053229.21-694411.0	& ab	& LMC-SC2-FIV	& 54197.04702	& ...	& ...	& 54197.05671	& 18.15	& 0.09 \\

\enddata
\label{tabrawobs}
\end{deluxetable}

\clearpage
\begin{deluxetable}{c c c c c c c c c}
\tablewidth{0pc}
\tablecaption{Averaged and mean magnitudes based on K-band template light-curves of RR Lyrae stars.}
\tablehead{\colhead{Star} & \colhead{Star} & \colhead{Period} &
  \colhead{J} & \colhead{$\sigma$} & \colhead{K} & \colhead{$\sigma$}
  & \colhead{$\langle K \rangle$} & \colhead{$\sigma$} \\
\colhead{ID} & \colhead{type} & \colhead{[days]} & \colhead{} & \colhead{} & \colhead{} & \colhead{} & \colhead{} & \colhead{} }
\startdata
OGLE051817.95-690358.9	& ab	& 0.4270819	& 18.81	& 0.08	& 18.55	& 0.13	& 18.47	& 0.08\\
OGLE051818.33-690513.2	& c	& 0.3543007	& 18.29	& 0.08	& 18.16	& 0.09	& 18.16	& 0.02\\
OGLE051818.71-690506.2	& ab	& 0.6248463	& 17.67	& 0.05	& 17.36	& 0.06	& 17.43	& 0.05\\
OGLE051820.31-690819.4	& c	& 0.2892113	& 18.12	& 0.08	& 18.06	& 0.10	& 18.07	& 0.04\\
OGLE051824.42-690459.4	& ab	& 0.6464268	& 18.28	& 0.06	& 17.95	& 0.07	& 17.99	& 0.09\\
OGLE051834.72-690550.6	& ab	& 0.4991066	& 18.20	& 0.06	& 17.99	& 0.08	& 18.07	& 0.02\\
OGLE051835.56-690406.0	& ab	& 0.4736237	& 18.22	& 0.08	& 18.18	& 0.09	& 18.19	& 0.10\\
OGLE051836.59-690839.6	& ab	& 0.5336746	& 18.55	& 0.09	& 18.47	& 0.05	& 18.38	& 0.05\\
OGLE051836.68-690659.7	& ab	& 0.5913999	& 18.27	& 0.07	& 18.10	& 0.04	& 18.03	& 0.04\\
OGLE051837.86-690821.3	& c	& 0.2873557	& 18.79	& 0.09	& 18.57	& 0.12	& 18.57	& 0.00\\
OGLE051841.62-690609.4	& ab	& 0.7517683	& 18.27	& 0.05	& 17.84	& 0.07	& 17.81	& 0.01\\
OGLE051842.24-690724.5	& ab	& 0.6597547	& 18.28	& 0.03	& 17.94	& 0.03	& 17.94	& 0.03\\
OGLE051843.33-690502.6	& ab	& 0.5676291	& 18.37	& 0.07	& 18.11	& 0.09	& 18.13	& 0.02\\
OGLE051853.07-690819.5	& ab	& 0.5805700	& 18.30	& 0.04	& 17.93	& 0.06	& 18.00	& 0.06\\
OGLE051923.06-693859.0	& ab	& 0.6438395	& 18.36	& 0.07	& 17.79	& 0.10	& 17.84	& 0.01\\
OGLE052301.40-700731.0	& c	& 0.2964723	& 18.58	& 0.05	& 18.39	& 0.05	& 18.39	& 0.05\\
OGLE052308.89-700746.7	& c	& 0.3425135	& 18.48	& 0.05	& 18.24	& 0.05	& 18.24	& 0.05\\
OGLE052317.13-700607.2	& ab	& 0.6020069	& 18.33	& 0.08	& 18.10	& 0.08	& 18.05	& 0.08\\
OGLE052318.43-700651.5	& ab	& 0.5521760	& 18.33	& 0.05	& 17.96	& 0.04	& 18.00	& 0.04\\
OGLE052321.67-700739.9	& ab	& 0.5214616	& 18.41	& 0.06	& 18.35	& 0.04	& 18.24	& 0.04\\
OGLE052328.70-700523.8	& c	& 0.3728943	& 17.96	& 0.09	& 17.78	& 0.10	& 17.77	& 0.05\\
OGLE052329.54-700713.1	& ab	& 0.6590143	& 18.06	& 0.09	& 17.71	& 0.06	& 17.75	& 0.04\\
OGLE052331.40-700529.1	& ab	& 0.6131165	& 18.02	& 0.09	& 17.78	& 0.08	& 17.85	& 0.05\\
OGLE052332.28-700654.1	& ab	& 0.6235510	& 17.36	& 0.06	& 17.28	& 0.06	& 17.31	& 0.00\\
OGLE052335.14-700505.4	& ab	& 0.5072454	& 18.30	& 0.07	& 18.10	& 0.09	& 18.07	& 0.01\\
OGLE052340.40-700639.8	& ab	& 0.5148516	& 18.04	& 0.07	& 18.07	& 0.08	& 18.10	& 0.08\\
OGLE052341.81-700632.4	& ab	& 0.5285457	& 18.84	& 0.12	& 18.56	& 0.12	& 18.64	& 0.20\\
OGLE052345.86-700504.2	& ab	& 0.5711875	& 18.10	& 0.07	& 17.95	& 0.09	& 17.88	& 0.05\\
OGLE052349.66-700327.1	& ab	& 0.6631955	& 18.17	& 0.12	& 18.03	& 0.12	& 18.04	& 0.11\\
OGLE052350.37-700443.2	& ab	& 0.5054740	& 18.53	& 0.10	& 18.24	& 0.11	& 18.26	& 0.04\\
OGLE052353.56-700432.4	& ab	& 0.5520142	& 18.31	& 0.07	& 18.13	& 0.08	& 18.21	& 0.13\\
OGLE052355.34-700802.6	& c	& 0.2628732	& 18.77	& 0.11	& 18.67	& 0.19	& 18.56	& 0.14\\
OGLE052357.93-700322.9	& ab	& 0.5125914	& 18.80	& 0.08	& 18.54	& 0.18	& 18.47	& 0.06\\
OGLE052400.52-700520.9	& ab	& 0.5816737	& 18.40	& 0.06	& 18.05	& 0.08	& 18.04	& 0.02\\
OGLE052405.11-700611.5	& ab	& 0.5950911	& 18.23	& 0.07	& 17.96	& 0.08	& 17.93	& 0.05\\
OGLE052418.64-700307.7	& c	& 0.3354890	& 18.28	& 0.11	& 18.18	& 0.16	& 18.18	& 0.16\\
OGLE052419.91-700616.7	& ab	& 0.5036958	& ...	& ...	& 17.90	& ...	& 17.83	& ...\\
OGLE052423.93-700736.4	& ab	& 0.5816925	& 18.32	& 0.05	& 18.12	& 0.04	& 17.93	& 0.04\\
OGLE052859.00-700822.2	& ab	& 0.5515840	& 18.97	& 0.10	& 18.31	& 0.13	& 18.08	& 0.13\\
OGLE052909.42-700828.1	& ab	& 0.9210850	& 17.70	& 0.05	& 17.42	& 0.07	& 17.31	& 0.07\\
OGLE052911.54-700836.5	& ab	& 0.5469046	& 18.58	& 0.06	& 17.99	& 0.07	& 17.99	& 0.07\\
OGLE052917.43-700734.5	& ab	& 0.6015387	& 18.34	& 0.04	& 18.06	& 0.07	& 18.12	& 0.07\\
OGLE052917.77-700535.6	& ab	& 0.7286710	& 18.43	& 0.07	& 17.81	& 0.07	& 17.68	& 0.07\\
OGLE052917.87-700841.1	& ab	& 0.4802590	& 18.54	& 0.05	& 18.23	& 0.08	& 18.15	& 0.08\\
OGLE052917.92-700712.2	& c	& 0.3383890	& 18.51	& 0.07	& 18.06	& 0.10	& 18.06	& 0.10\\
OGLE052929.39-700551.0	& ab	& 0.6215635	& 18.00	& 0.04	& 17.59	& 0.06	& 17.65	& 0.06\\
OGLE052932.88-700842.3	& ab	& 0.5442339	& 18.37	& 0.06	& 18.26	& 0.09	& 18.27	& 0.09\\
OGLE052933.54-700715.9	& ab	& 0.5340639	& 18.58	& 0.05	& 18.03	& 0.08	& 17.99	& 0.08\\
OGLE052936.94-700721.1	& ab	& 0.5858587	& 18.37	& 0.05	& 18.35	& 0.08	& 18.34	& 0.08\\
OGLE052937.91-700844.9	& c	& 0.3758092	& 18.24	& 0.05	& 17.95	& 0.08	& 17.95	& 0.08\\
OGLE052949.93-700558.2	& ab	& 0.5625618	& 18.66	& 0.06	& 18.22	& 0.09	& 18.17	& 0.09\\
OGLE053147.30-694349.6	& ab	& 0.5074034	& 18.11	& 0.05	& 17.82	& 0.07	& 17.90	& 0.07\\
OGLE053147.82-694142.2	& ab	& 0.6455259	& 18.05	& 0.05	& 17.81	& 0.07	& 17.70	& 0.07\\
OGLE053155.92-694232.4	& ab	& 0.5211680	& 18.35	& 0.05	& 18.19	& 0.08	& 18.17	& 0.08\\
OGLE053200.94-694219.1	& ab	& 0.6062906	& 18.20	& 0.09	& 18.01	& 0.12	& 17.95	& 0.12\\
OGLE053203.62-694210.7	& ab	& 0.4796139	& 18.26	& 0.04	& 17.99	& 0.07	& 18.05	& 0.07\\
OGLE053206.72-694342.1	& ab	& 0.5192097	& 18.51	& 0.04	& 18.38	& 0.09	& 18.42	& 0.09\\
OGLE053206.78-694224.7	& ab	& 0.5230170	& 17.98	& 0.04	& 18.02	& 0.07	& 18.04	& 0.07\\
OGLE053207.32-694117.2	& ab	& 0.5626089	& 17.76	& 0.03	& 17.53	& 0.06	& 17.58	& 0.06\\
OGLE053207.88-694058.7	& ab	& 0.6218200	& 18.15	& 0.04	& 17.93	& 0.07	& 18.02	& 0.07\\
OGLE053217.55-694317.6	& ab	& 0.5485307	& 17.95	& 0.05	& 17.87	& 0.07	& 17.87	& 0.07\\
OGLE053219.15-694215.2	& ab	& 0.6521316	& 18.22	& 0.05	& 17.94	& 0.07	& 17.97	& 0.07\\
OGLE053222.39-693952.9	& ab	& 0.6769293	& 18.23	& 0.07	& 18.03	& 0.11	& 18.10	& 0.11\\
OGLE053225.42-694438.2	& ab	& 0.6427599	& 18.38	& 0.07	& 18.03	& 0.12	& 18.07	& 0.12\\
OGLE053229.21-694411.0	& ab	& 0.5960780	& ...	& ...	& 18.15	& 0.09	& 18.10	& 0.09\\

\enddata
\label{tabavobs}
\end{deluxetable}

\clearpage
\begin{deluxetable}{c c c c c}
\tablewidth{0pc}
\tablecaption{PL relations determined for averaged data}
\tablehead{ \colhead {Data set} & \colhead {J slope} & \colhead {J zeropoint} & \colhead {K slope} & \colhead {K zeropoint} }
\startdata
RRab+RRc\\
Free fit   &     -1.381$\pm$0.468 &  17.953$\pm$0.123     &      -2.192$\pm$0.399  & 17.493$\pm$0.108   \\
Sollima et al. (2008)   &     ...             &  ...                 &      -2.380         & 17.443$\pm$0.026   \\
Bono   et al. (2003)    &     ...             &  ...                 &      -2.101         & 17.516$\pm$0.026   \\
Catelan  et al. (2004)  &     -1.773        &  17.850$\pm$0.030     &      -2.353         & 17.450$\pm$0.026   \\
\\
RRab\\
Sollima  et al. (2008)  &    ...              & ...                 &      -2.380         & 17.463$\pm$0.029   \\
Bono     et al. (2003)  &    ...              & ...                 &      -2.101         & 17.531$\pm$0.029   \\
Catelan  et al. (2004)  &    -1.773         & 17.869$\pm$0.033     &      -2.353         & 17.469$\pm$0.029   \\
\enddata
\label{tabplav}
\end{deluxetable}

\clearpage
\begin{deluxetable}{c c c}
\tablewidth{0pc}
\tablecaption{PL relations determined for mean data, using the light-curve templates}
\tablehead{ \colhead {Data set} & \colhead {K slope} & \colhead {K zeropoint}}
\startdata
RRab+RRc\\
Sollima et al. (2008)  &     -2.380   &       17.421$\pm$0.026   \\
Bono    et al. (2003)  &     -2.101   &       17.500$\pm$0.026   \\
Catelan et al. (2004)  &     -2.353   &       17.434$\pm$0.026   \\
\\
RRab\\
Sollima et al. (2008)  &     -2.380    &      17.445$\pm$0.029   \\
Bono    et al. (2003)  &     -2.101    &      17.512$\pm$0.029   \\
Catelan et al. (2004)  &     -2.353    &      17.451$\pm$0.029   \\
\enddata
\label{tabplmean}
\end{deluxetable}

\clearpage
\begin{deluxetable}{c c c c}
\tablewidth{0pc}
\tablecaption{Determined true distance moduli for different data sets and PL relations.}
\tablehead{ \colhead {Data set} & \colhead {Sollima et al. (2008)} & \colhead
  {Bono et al. (2003)} & \colhead {Catelan et al. (2004)}}
\startdata
RRab+RRc (J)	&	...		&	...		&	18.546 $\pm$ 0.026 \\
RRab+RRc (K)	&	18.580 $\pm$ 0.026	&	18.621 $\pm$ 0.026	&	18.607 $\pm$ 0.026 \\
RRab (J)	&	...		&	...		&	18.565 $\pm$ 0.029 \\
RRab (K)	&	18.600 $\pm$ 0.029	&	18.620 $\pm$ 0.029	&	18.611 $\pm$ 0.029 \\
RRab+RRc ($\langle K\rangle$)	&	18.564 $\pm$ 0.026	&	18.604 $\pm$ 0.026	&	18.590 $\pm$ 0.026 \\
RRab ($\langle K\rangle$)	&	18.582 $\pm$ 0.029	&	18.617 $\pm$ 0.029 & 18.608 $\pm$ 0.029 \\
\enddata
\label{tabdist}
\end{deluxetable}

\end{document}